\def \SAIT #1 #2 {{\em Mem.\ Soc.\ Astron.\ It.\/} {\bf #1}, #2}
\def \MESS #1 #2 {{\em The Messenger\/} {\bf #1}, #2}
\def \ASTRNACH #1 #2 {{\em Astron. Nach.\/} {\bf #1}, #2}
\def \AAP #1 #2 {{\em Astron. Astrophys.\/} {\bf #1}, #2}
\def \AAL #1 #2 {{\em Astron. Astrophys. Lett.\/} {\bf #1}, L#2}
\def \AAR #1 #2 {{\em Astron. Astrophys. Rev.\/} {\bf #1}, #2}
\def \AAS #1 #2 {{\em Astron. Astrophys. Suppl. Ser.\/} {\bf #1}, #2}
\def \AJ #1 #2 {{\em Astron. J.\/} {\bf #1}, #2}
\def \ANNREV #1 #2 {{\em Ann. Rev. Astron. Astrophys.\/} {\bf #1}, #2}
\def \APJ #1 #2 {{\em Astrophys. J.\/} {\bf #1}, #2}
\def \APJL #1 #2 {{\em Astrophys.. J. Lett.\/} {\bf #1}, L#2}
\def \APJS #1 #2 {{\em Astrophys. J. Suppl.\/} {\bf #1}, #2}
\def \APSS #1 #2 {{\em Astrophys. Space Sci.\/} {\bf #1}, #2}
\def \ASR #1 #2 {{\em Adv. Space Res.\/} {\bf #1}, #2}
\def \BAIC #1 #2 {{\em Bull. Astron. Inst. Czechosl.\/} {\bf #1}, #2}
\def \JSQRT #1 #2 {{\em J. Quant. Spectrosc. Radiat. Transfer\/} {\bf #1}, #2}
\def \MN #1 #2 {{\em Mon. Not. R. Astr. Soc.\/} {\bf #1}, #2}
\def \MEM #1 #2 {{\em Mem. R. Astr. Soc.\/} {\bf #1}, #2}
\def \PLR #1 #2 {{\em Phys. Lett. Rev.\/} {\bf #1}, #2}
\def \PASJ #1 #2 {{\em Publ. Astron. Soc. Japan\/} {\bf #1}, #2}
\def \PASP #1 #2 {{\em Publ. Astr. Soc. Pacific\/} {\bf #1}, #2}
\def \NAT #1 #2 {{\em Nature\/} {\bf #1}, #2}
\def\ltsima{$\; \buildrel < \over \sim \;$}
\def\simlt{\lower.5ex\hbox{\ltsima}}

\documentstyle{memsait}
\input epsf.sty
\begin{opening}
\title{Clusters of Galaxies - Keys to Cosmology} 
\author{Sabine Schindler}

\institute{Astrophysics Research Institute, Liverpool John Moores University,
Twelve Quays House, Birkenhead CH41 1LD, U.K.}

\date{} 
\end{opening}

\begin{document}

\oddpagefooter{}{}{} 
\evenpagefooter{}{}{} 
\ 
\bigskip

\begin{abstract}
Various cosmological applications of galaxy clusters are
presented. Clusters are used to determine the baryon fraction, dark
matter distribution and the matter density $\Omega_m$ of the
universe. They also contain a wealth of information about structure
formation and evolution on different scales: large-scale structure, cluster
formation and also galaxy formation 
via the interaction of cluster galaxies with the intra-cluster
medium.  In particular, 
the X-ray satellites CHANDRA and XMM yield exciting
new results for galaxy cluster physics and cosmology. 
\end{abstract}

\section{Introduction}

Clusters of galaxies are used for various kinds of
cosmological studies. Clusters can be observed out to large
distances and hence they can trace the distribution of matter on large
scales. Furthermore, a comparison of the properties of these distant
clusters with that of nearby clusters reveals the evolution that clusters
undergo between a redshift of 0.5-1 and now. Both large-scale
structure and cluster evolution depend sensitively on cosmological
parameters. Another interesting point is that clusters are closed
systems: no matter can leave
the deep potential well. Therefore all the metals that have been
processed inside a cluster must still be present, i.e. all the traces
of the
formation process of the cluster and its galaxies are still observable.
Clusters -- the largest bound structures in the universe --
for many applications can be regarded as being representative of the
universe as a whole, because they accumulate matter from a
relatively large
volume (a few tens of Mpc). Moreover, the crossing time (= the time it
takes a galaxy to move from one end of the cluster to the other) is
not much smaller than a Hubble time. That means, not all of the
traces of the formation process are wiped out, but some information about
the early universe is still present.

In the following we will present selected topics which are very
promising for cosmology: the 
distribution of dark and baryonic matter, and a brief overview of
various ways to study structure formation and the evolution of
different cluster components by investigating their interaction.  
Throughout this article we use $\rm{H}_0 = 50$ km/s/Mpc.

\begin{figure}
\epsfxsize=10.2cm 
\hspace{1cm}\epsfbox{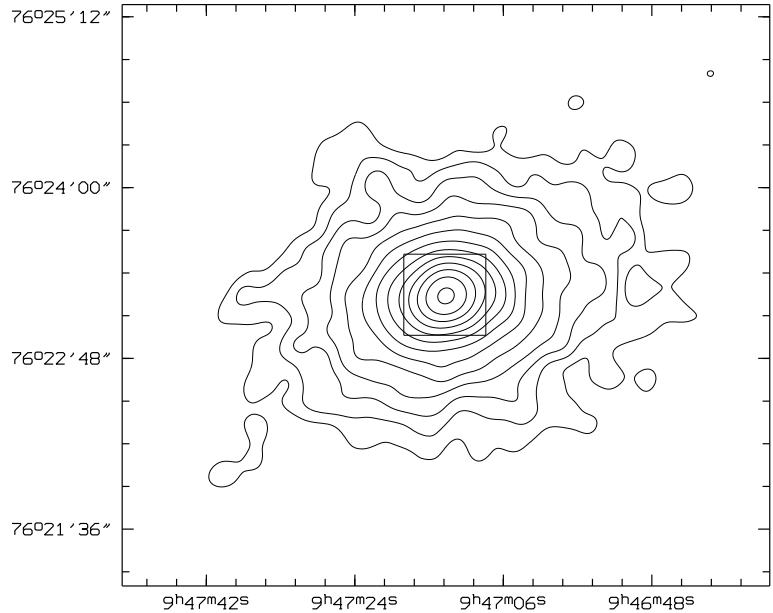} 
\vskip.2in
\caption
{CHANDRA image of the cluster RBS797. The X-ray emission from the hot
gas in the cluster is relatively
regular with an ellipticity of 1.3-1.4 in E-W direction.
The size of the image is 4.7 arcmin on the side.
The central square marks the region shown with higher resolution 
in Fig.~\ref{fig:rbs797_cen}.
}
\label{fig:rbs797}
\end{figure}

\section{Distribution of Dark and Baryonic Matter}

The deep potential wells of galaxy
clusters have accumulated matter from a large volume. Hence the ratio
of baryonic to total matter in clusters is 
representative of the universe as a whole if all the matter is
accumulated indiscriminately. Most of the baryons are in the
intra-cluster medium (ICM) -- the hot gas that fills all the space
between the galaxies (see Fig.~1). 
Typical values for the gas mass fraction are
10-25\%, while the mass fraction in galaxies is only 3-5\% of the
total cluster mass.  
Measuring distributions and ratios of baryonic and non-baryonic matter
in galaxy clusters is particularly interesting because there are
several independent methods to determine the total cluster mass. In
this article I will concentrate  on the mass determination by
X-ray observations. Assuming that the gas traces the potential well (=
hydrostatic equilibrium) the total mass can be measured through the
X-ray surface brightness distribution and the gas temperature.  

This approach is correct only if the magnetic pressure is negligible
compared to the thermal pressure. Otherwise the mass is
underestimated. To test the 
influence of the magnetic fields on the mass determination we used
magneto-hydrodynamic simulations by Dolag et al. (1999). We found that
the magnetic field causes only an underestimation of the total mass in
the very centre of relaxed clusters and even there it is only a very
small effect of typically not more than 5\% (Dolag \& Schindler
2000). However, if the clusters are not relaxed, but in the process of
merging, the mass can be underestimated considerably. Therefore the
mass determination should be either restricted to virialised clusters
or be performed very cautiously in clusters which are not in equilibrium.

Several groups determined
gas mass fractions from X-ray observations in samples of nearby and 
distant clusters, e.g. Mohr et al. (1999): $f_{gas}=0.21$,
Ettori \& Fabian (1999): $f_{gas}=0.17$,
Arnaud \& Evrard (1999): $f_{gas}=0.16 - 0.20$,
Schindler (1999):  $f_{gas}=0.18$ (see also Fig.~2).
All these determinations depend on the radius where the mass fraction
is determined, because the gas mass fraction increases slightly 
with radius. Therefore the gas mass fractions have to be calculated
for equivalent radii.
In the above mentioned analyses the mass was determined
within a radius $r_{500}$ from the cluster centre. 
This radius encompasses a volume 
that has a density of 500 $\times$ the
critical density of the universe $\rho_{crit}$. 
Out to this radius the X-ray profiles
necessary for the analysis could be measured reliably.

\begin{figure}
\epsfxsize=8.2cm 
\hspace{2cm}\epsfbox{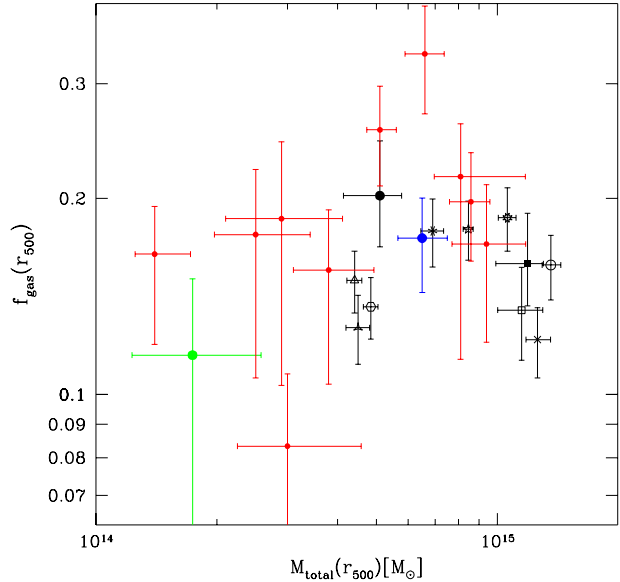} 
\vskip.2in
\caption[h]
{Gas mass fraction in galaxy clusters versus total mass of the
cluster
(from Castillo-Morales \& Schindler, in prep.). 
}
\label{fig:fgas_Mt}
\end{figure}

To determine $\Omega_m$, the gas mass fraction $f_{gas}$ must
be compared to the baryon density in the universe
$\Omega_B$. Burles \& Tytler (1998a,b) found $\Omega_B \simlt 0.08$  
from primordial nucleosynthesis. The ratio of the baryon density and
the gas mass fraction 
yields an upper limit for the matter density 
$\Omega_m < {\Omega_B\over f_{gas}} \approx 0.3 - 0.4$.

The baryon fraction can also be determined in a different way: 
measurements of  the Sunyaev-Zel'dovich effect -- inverse-Compton
scattering of the Cosmic Microwave Background photons by the
hot intra-cluster gas -- shifts the CMBR spectrum to slightly higher energies. 
As this effect is proportional to the gas density,
the density profile can be determined directly. 
Only an additional measurement
of the gas temperature is necessary from X-rays. 
The mean gas mass fraction found by
Grego et al. (2001) in a sample of 18 clusters ($f_{gas}=0.16$) 
is very similar to the X-ray results.
Hence they derive also a similar upper limit for the matter density
$ \Omega_m < 0.4$.
In these analyses only the mass in the intra-cluster gas was taken into
account. Baryons in the galaxies were neglected. If they were to be
included, 
the baryon fraction would increase  slightly and hence ever more stringent
constraints on $\Omega_m$ could be placed.

\begin{figure}
\epsfxsize=8.2cm 
\hspace{2cm}\epsfbox{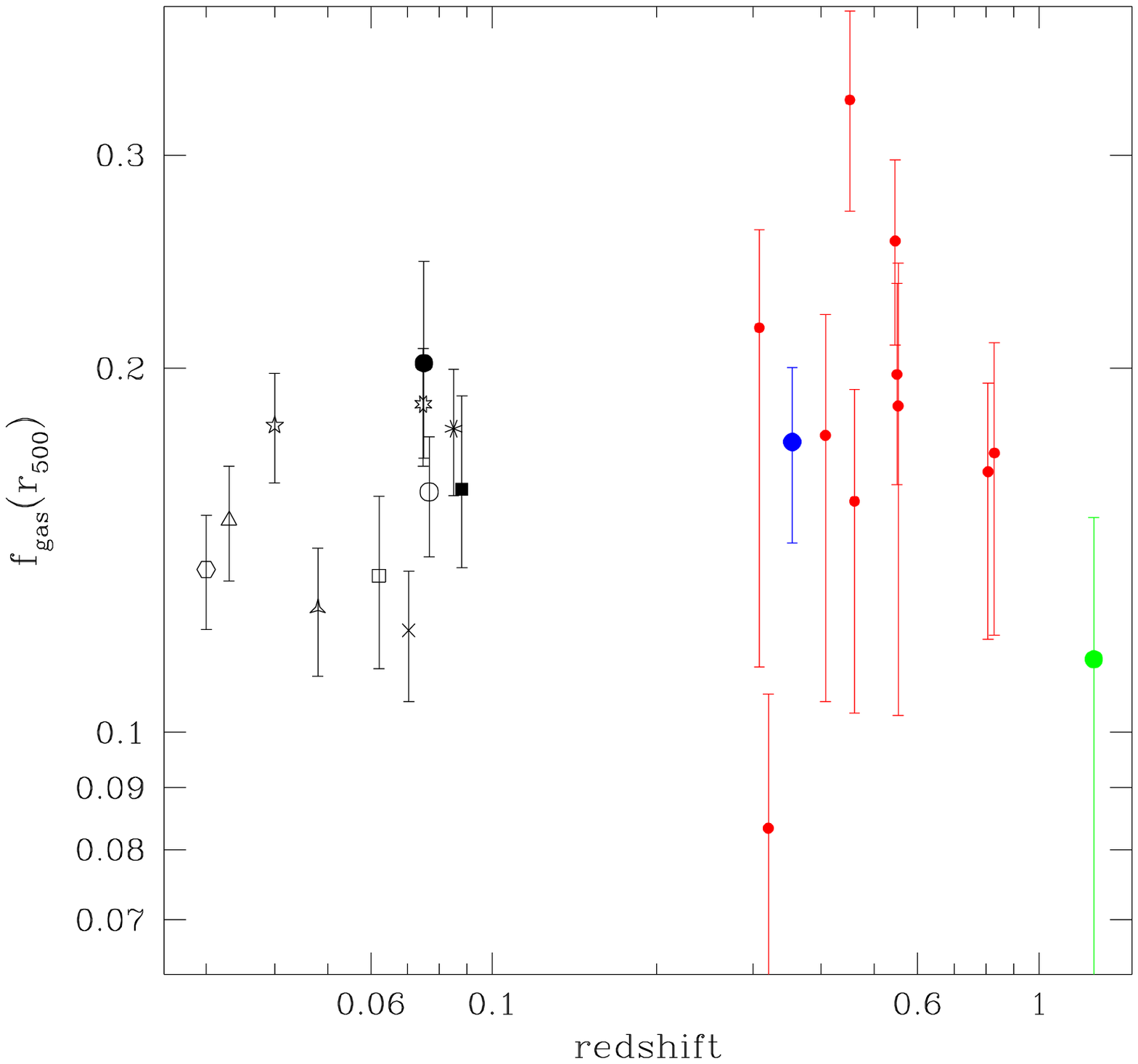} 
\vskip.2in
\caption{Gas mass fraction in galaxy clusters versus redshift. The
points at z=0.35 and z=1.3 come from CHANDRA 
observations, all the others use a combination of ROSAT and ASCA data 
(from Castillo-Morales \& Schindler, in prep.). 
}
\label{fgas_z}
\end{figure}

For cosmology and cluster formation it is very interesting
to know how the baryon fraction in clusters changes with time. In
Fig.~\ref{fgas_z} the gas mass fraction is plotted versus the redshift. It is
clear that with the previous ROSAT and ASCA observations it was very
hard to see any evolution due to the large observational 
uncertainties. With XMM and
CHANDRA we will be able to measure gas mass fractions out to redshifts
of almost unity with good accuracy. 

Even with the old observations one can 
see clearly that the gas fraction varies considerably from cluster to
cluster. These different baryon fractions reflect probably the very
early distribution of baryonic and non-baryonic matter and they are
therefore of high interest for cosmology. 

\section{Cosmological Structure Formation}

Clusters of galaxies can be used to study how structures form on large
scales. The formation and evolution of clusters depends very
sensitively on cosmological parameters like the mean matter density in
the universe $\Omega_m$ (Thomas et al. 1998; Jenkins et al. 1998;
Beisbart et al. 2001).
Therefore it is of great importance to determine the dynamical state
of clusters at different redshifts, i.e. at different evolutionary
states. 
 
There are different approaches to determine the dynamical state. X-ray
observations yield a projected image of the square of the ICM density
and hence they often give a first idea. A much more
detailed picture can be obtained through the temperature distribution
of the ICM, because shocks and gas compressions due to mergers are
visible clearly in temperature maps. The excellent capabilities for
spatially resolved spectroscopy of the new X-ray observatories XMM and
CHANDRA finally make a detailed temperature analysis possible, see
e.g. the temperature maps by Arnaud et al. (2001), Neumann et
al. (2001), Markevitch et al. (2000). 

Optical observations of the clusters galaxies are also very useful for
the determination of the evolutionary state, because both the spatial
distribution and the velocity distribution of the cluster galaxies 
are good indicators.  
 
In the shocks produced during the merging process particles can be
accelerated to relativistic energies. These relativistic 
particles and hence a
previous merger event can be detected in two ways. (1) Inverse Compton
scattering of photons from the Cosmic Microwave Background Radiation
by these relativistic particles can cause a hard X-ray excess in the
spectrum, which has been discovered in a few clusters (Fusco-Femiano
et al. 1999, 2000, 2001). (2) Due to the presence of magnetic fields,
synchrotron radiation is emitted in the form of radio halos
(Markevitch \& Vikhlinin 2001, Feretti et al. 2001). For a detailed
analysis it is necessary to know the distribution of the magnetic
field within the cluster. In a recent analysis we found that the
magnetic field decreases with radius almost in the same way as the ICM
density decreases with radius (Dolag et al. 2001).

Hence, the optimal approach for the determination of the dynamical
state is a combination of many
wavelengths -- X-ray, optical, radio and hard X-ray. 

\section{Interaction of Galaxies and the Intra-Cluster Gas}

The various components in a cluster interact with each other. In
particular the interaction between the cluster galaxies and the ICM is
very important. Hence this interaction is 
more and more studied at the different
wavelengths. Not only the energy budget and entropy of the ICM are
influenced but also the ICM content of heavy elements. The amount of
iron in the ICM is e.g. of the same order of magnitude as that in 
the cluster galaxies (Mushotzky 1999). 
The heavy elements in the ICM cannot be of primordial origin, but must
have been produced in the cluster galaxies and subsequently been
transported into the ICM. Several transport processes are possible:
ram-pressure stripping, galactic winds, galaxy-galaxy interaction, and
jets from active galaxies. So far very controversial results have been
obtained in various studies which tried to determine for example
the dominant metal enrichment process at a given evolutionary
state (see e.g. Metzler \& Evrard 1994; Gnedin 1998; Murakami \& Babul
1999, Aguirre et al. 2001). 
If we can find the correct answers, we will learn a lot about
galaxy formation and cluster formation.  

\begin{figure}
\epsfxsize=5.2cm 
\hspace{3.5cm}\epsfbox{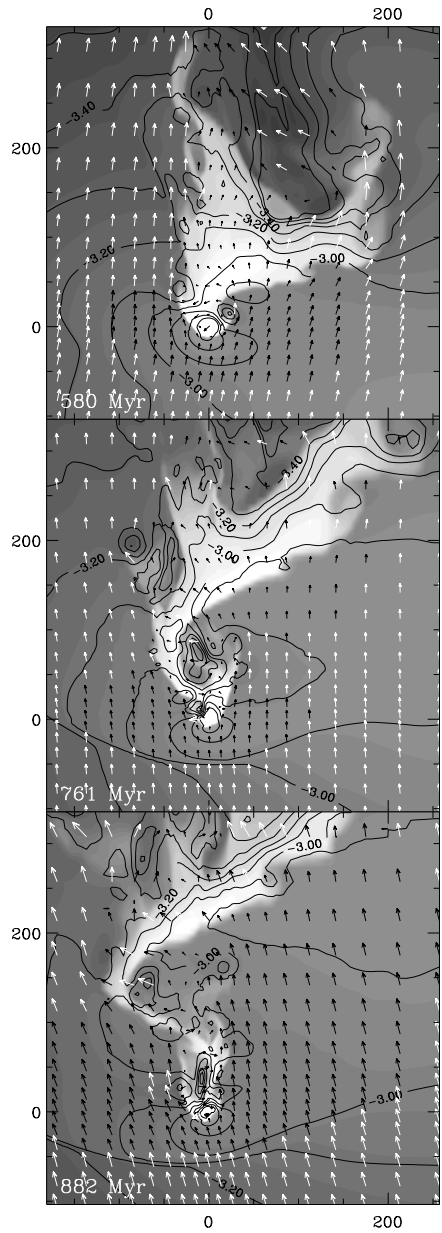} 
\vskip.2in
\caption{Gas density (grey scale) and pressure (contours) of a galaxy
moving downwards towards the cluster centre. The arrows show the Mach
vectors (white when $M>1$, black otherwise). The gas of the galaxy is
stripped due to ram pressure (from Toniazzo \& Schindler 2001). 
}
\label{fig:toniazzo1}
\end{figure}

The new X-ray satellites XMM and CHANDRA revolutionise also this
field. It is now possible to measure reliable metallicities out to a
redshift of about unity, which was hardly possible before with
e.g. ASCA measurements (Schindler 1999). Furthermore, the metal
distribution within the clusters can be measured with much higher
accuracy than up to now with e.g. BeppoSAX observations (De Grandi \&
Molendi 2001). Also different chemical elements can be distinguished now.
In order to interpret these excellent new data
correctly several numerical simulations have been performed to study
the different enrichment processes, for example, ram-pressure
stripping of galaxies when they move through the ICM (Abadi et
al. 1999, Quilis et al. 2000, Mori \& Burkert 2000, Schulz \& Struck
2001; Toniazzo \& Schindler 2001; see also Fig.~\ref{fig:toniazzo1}). 
Comprehensive simulations on cluster scales will give us new insights
into the various metal enrichment processes of the ICM.


\begin{figure}
\epsfxsize=8.2cm 
\hspace{2.5cm}\epsfbox{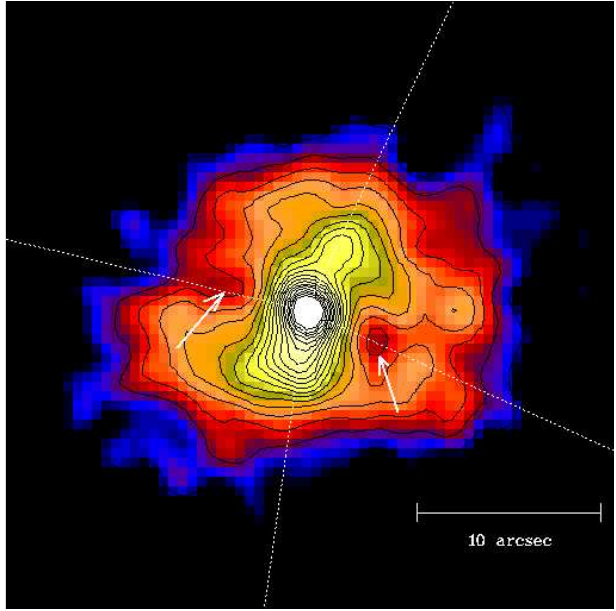} 
\vskip.2in
\caption{X-ray image of the central (32 arcsec)$^2$ of the
cluster RBS797. There are clear minima in the X-ray emission 
at about 5 arcseconds from the cluster centre (see
arrows). The minima are opposite to each other with respect to the cluster
centre. 
An excess of emission is found in perpendicular directions. The dotted
lines mark the traces shown in Fig.~\ref{fig:traces}.
}
\label{fig:rbs797_cen}
\end{figure}

\begin{figure}
\epsfxsize=6.8cm 
\hspace{2cm}\epsfbox{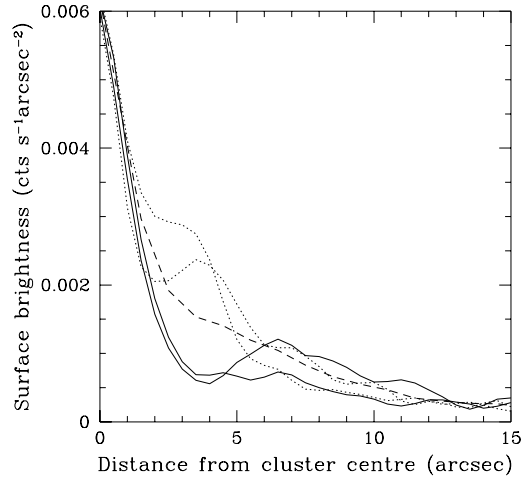} 
\vskip.2in
\caption{Dependence of the surface brightness in RBS797 on direction: 
the X-ray emission from the centre in direction
of the minima (solid lines, position angles $77^{\circ}$ and
$243^{\circ}$, N over E)  
and in direction of the maxima
(dotted lines, position angles $171^{\circ}$ and
$337^{\circ}$). 
There is an X-ray deficit of a factor of 3 - 4
in the ``holes'' compared to the perpendicular directions.  
The dashed line shows an average profile integrated over all angles.
}
\label{fig:traces}
\end{figure}

\section{Selected Highlights from XMM and CHANDRA observations of Galaxy Clusters}

Observations of galaxy clusters with the new X-ray satellites XMM and
CHANDRA have revealed several surprising results. The high spatial
resolution of CHANDRA enables us to look in detail at the cluster
centres. In some clusters ``depressions'' in the X-ray emission have
been found, e.g.   
RBS797 (Schindler et al. 2001, see also Fig.~\ref{fig:rbs797_cen} and
Fig.~\ref{fig:traces}), 
Perseus cluster (Fabian et al. 2000), and 
Hydra-A (McNamara et al. 2000). In these clusters the intra-cluster 
gas has been pushed from the 
low X-ray emission regions to the high X-ray emission regions by the
pressure of relativistic particles in radio jets, which come from
the active galaxy in the cluster centre. These clusters are
hence ideal
objects to investigate the interaction of jets with the hot 
intra-cluster medium. 

In several CHANDRA observations surprising ``edges'' have been
discovered, e.g. in A3667 
(Vikhlinin et al. 2001). These jumps in the X-ray surfaces brightness are
not shock fronts, because the temperature increases while the 
surface brightness
drops, i.e. the two sides are in pressure equilibrium. Therefore they
were named ``cold fronts''. They have a width of about 5kpc, which is 2-3
times smaller than the Coulomb mean free path. It is likely that
the transport processes here are suppressed by magnetic fields. 

The XMM spectra of cooling flow regions in the centres of some clusters
revealed that cool gas below a temperature of 1.5keV is absent,
e.g. in A1835 
(Peterson et al. 2001). This absence of cool gas was very surprising
as it provides a serious problem to standard cooling flow models.

\section{Summary}

Clusters of galaxies are ideal diagnostic tools to determine
cosmological parameters in very different ways. In this article only a
small selection is presented. The determination of cluster masses and
baryon fractions in clusters of galaxies is one of the very active
fields at the moment. Also the formation process of clusters is
studied in different ways. 
Interaction of different cluster components (galaxies,
intra-cluster medium) is a topic of increasing interest, because
the interaction leaves traces in the metallicities which can be
used to infer details of the formation process of clusters and of
galaxies. In particular the new X-ray
telescopes XMM and CHANDRA yield at the moment extremely
exciting results which give new insights in cluster physics and
cosmology. These observations will soon answer many of the
open questions in cosmology.    

\acknowledgements 
I thank Betty De Filippis and Africa Castillo-Morales for their 
sedulous commitment 
and Phil James for carefully reading the manuscript.

\vskip 1cm
\noindent
{\bf DISCUSSION}
\vskip 0.4cm
\noindent
{\bf N. PANAGIA:} About the anti-correlation found between X-ray and
radio emission, I am wondering which is the cause and which the
effect: is it the radio structure that pushes the X-ray emitting gas
or, viceversa, is it the hot gas that confines the radio emitting matieral?

\vskip 0.3cm
\noindent
{\bf S. SCHINDLER:} 
In the current model the pressure of the relativistic particles in the
radio jets push away the X-ray emitting gas until the system is in pressure
equilibrium.  

\vskip 0.3cm
\noindent
{\bf G. KANBACH:} 
Radio observations show that cosmic ray pressure could influence the
intra-cluster medium. Is the cosmic ray pressure generally taken into
account in the dynamical balance of the intra-cluster medium?

\vskip 0.3cm
\noindent
{\bf S. SCHINDLER:} 
So far cosmic ray pressure is usually not taken into account. For
standard mass
determinations from X-ray observations
only the thermal pressure of the intra-cluster medium
is taken into account.

\vskip 0.3cm
\noindent
{\bf W. KUNDT:} 
What topology was assumed for the magnetic field simulations in
clusters?

\vskip 0.3cm
\noindent
{\bf S. SCHINDLER:} 
Different types of initial magnetic seed fields were used: completely
homogeneous and also chaotic initial
magnetic field structures. This field is amplified by 
compression during the cluster collapse. It was shown that the final
field structure is determined only by the 
dynamics of the cluster collapse and not by the inital conditions.

\end{document}